\documentstyle{elsart}
\begin{document}
\begin{frontmatter}

\title{ Production Asymmetry of $D_s$ from 600 GeV/c $\Sigma^-$ and 
$\pi^-$
beam}

\centerline{(SELEX Collaboration)}

\author[Iowa]{M.~Kaya\thanksref{tre}},
\author[PNPI]{G.~Alkhazov},
\author[PNPI]{A.G.~Atamantchouk\thanksref{tra}},
\author[ITEP]{M.Y.~Balatz\thanksref{tra}},
\author[PNPI]{N.F.~Bondar},
\author[Fermi]{P.S.~Cooper},
\author[Flint]{L.J.~Dauwe},
\author[ITEP]{G.V.~Davidenko},
\author[MPI]{U.~Dersch\thanksref{trb}},
\author[ITEP]{A.G.~Dolgolenko},
% ...
\address[Fermi]{Fermi National Accelerator Laboratory, Batavia, IL 60510, 
U.S.A.}
% ...
\address[ITEP]{Institute of Theoretical and Experimental Physics, Moscow, 
Russia}
\address[MPI]{Max-Planck-Institut f\"ur Kernphysik, 69117 Heidelberg, 
Germany}
% ...
\address[PNPI]{Petersburg Nuclear Physics Institute, St. Petersburg, 
Russia}
% ...
\address[Iowa]{University of Iowa, Iowa City, IA 52242, U.S.A.}
\address[Flint]{University of Michigan-Flint, Flint, MI 48502, U.S.A.}
% ...
\thanks[tra]{deceased}
\thanks[trb]{Present address: Infinion, M\"unchen, Germany}
\thanks[trc]{Present address: Imperial College, London SW7 2BZ, U.K.}
\thanks[trd]{Present address: Instituto de F\'{\i}sica da Universidade 
Estadual de Campinas, UNICAMP, SP, Brazil}
\thanks[tre]{Present address: Kafkas University, Kars, Turkey}
\thanks[trf]{Present address: Physik-Department, Technische Universit\"at 
M\"unchen, 85748 Garching, Germany}
\thanks[trg]{Present address: The Boston Consulting Group, M\"unchen, 
Germany}
\thanks[trh]{Present address: Instituto de F\'{\i}sica Te\'orica da 
Universidade Estadual Paulista, S\~ao Paulo, Brazil}
\thanks[tri]{Present address: Lucent Technologies, Naperville, IL}
\thanks[trj]{Present address: SPSS Inc., Chicago, IL}
\thanks[trk]{Present address: University of Alabama at Birmingham, 
Birmingham, AL 35294}
\thanks[trm]{Present address: DOE, Germantown, MD}
\thanks[trn]{Present address: Solidum, Ottawa, Ontario, Canada}
\thanks[tro]{ Present address: Siemens Medizintechnik, Erlangen, Germany}
\thanks[trp]{Present address: Deutsche Bank AG, Eschborn, Germany}

%
%

\begin{abstract}

The production of $D_s^-$ relative to $D_s^+$ as a function of
$x_F $ with 600 GeV/c $\Sigma^-$ beam is measured 
in the interval $0.15 < x_F < 0.7$ by the SELEX (E781)
experiment at Fermilab. The integrated charge asymmetries with 600 GeV/c
$\Sigma^-$ beam ($0.53\pm0.06$) and $\pi^-$ beam ($0.06\pm0.11$)  are 
also compared. The results show the $\Sigma^-$ beam fragments play a role 
in the
production of $D_s^-$, as suggested by the leading quark model.

\end{abstract}
\end{frontmatter}

\input{psfig}

%
%
\section{Introduction}

Perturbative Quantum Chromodynamics calculations in leading or
next-to-leading order predict very small or no asymmetry in the $x_F$ or 
$p_t$
distributions for charm and anticharm production\cite{PQCD1,PQCD2}.
However, fixed target data show some asymmetry between the
production of some charm and anticharm hadrons in hadron-hadron
interactions\cite{wa89,e791}. SELEX has shown strong, beam-dependent
asymmetries in $\Lambda_c^+$ production\cite{fernanda}. 
This experiment finds that $D_s^{\pm}$ production from a
$\Sigma^-$ beam (but not a $\pi^-$ beam) also has a large
production asymmetry.
This asymmetry could be due to the fact that the beam hadron
shares a quark with one of the charge states (hence leading particle) and
not with the other charge state (non-leading). This is sometimes called
``the leading particle effect." 

For a $\Sigma^-$($sdd$) beam the $D_s^-$($\overline{c} s$) shares an $s$
quark with the beam hadron and is a leading particle, whereas $D_s^+(c
\overline{s})$ is not.  For a $\pi^-$($\overline{u} d$) beam, neither 
$D_s^-$ nor $D_s^+$ is leading. 
Several theoretical models have been proposed to
explain charm hadroproduction in the framework of non-perturbative
hadronization.
Among the proposed models are the color-drag string model\cite{string},
which is pronounced at high $x_F$ and is independent of $p_t$, and the
intrinsic charm model\cite{intrinsic}, which manifests itself at low
$p_t$ and larger $x_F$.

%
%

\section{Apparatus }

Data were taken during the 1996-97 fixed target run at Fermilab.
The 600 GeV/c negative hadron beam used in this measurement was 
composed of approximately 50\% $\Sigma^-$ and 50\% $\pi^-$. Beam particles 
were 
tagged with a transition radiation detector system located
upstream of the charm production target.

A segmented target consisting of two copper sheets and three diamond
sheets, each spaced by 1.5 cm, was used to produce charm particles.
The total target thickness was $4.2\%$ of an interaction length for 
protons.

The SELEX experiment used a three-stage spectrometer 
designed for large acceptance at $x_F >0.1$ and for detecting the decay 
of charm particles. Each stage included a bending magnet and a detector 
system. 

SELEX used an online trigger to identify charm particles. The hardware 
trigger required at least 4 charged hadrons in the forward 150 mrad cone 
and 2 hits from positive track candidates in a hodoscope after second 
magnet (M2). The software trigger made a full vertex reconstruction of the 
beam track and all tracks in the M2 spectrometer. Trigger conditions were 
also included in the simulation. 

The RICH detector, located after the second spectrometer system, was 
filled 
with neon gas at room temperature and 1.05 atm
pressure. It identified charged hadrons whose trajectories went through 
the fiducial volume, typically requiring $p>$ 23 GeV/c\cite{richpaper}. Full reconstruction of 
the secondary vertex was provided by linking RICH-identified tracks back 
through 
the second stage magnetic spectrometer to the vertex silicon
detector and associating them at a common decay vertex downstream of 
the primary interaction vertex. 

%
%
%
%
\section{Data Analysis}

Initial data selection required two kaons to be identified by the
RICH (ratio of the likelihoods: ${\cal L}(K)/{\cal L}(\pi)\ge1$) and the
primary vertex to be in a target sheet.

The criteria used to select $D_s$ candidates included the following cuts:

i) Secondary vertex separation significance $L/\sigma > 9$ where $L$ is 
the longitudinal separation between primary and secondary vertex and 
$\sigma$ is the error on $L$.

ii) Secondary vertex was at least 100 microns outside of the target material.

iii) Each secondary track was extrapolated back to primary vertex 
$z$-position to evaluate the transverse miss distance $b$. The second
largest miss distance had to have $(b/\sigma_b)^2 > 8$, where  $\sigma_b$
is the error on $b$.

These cuts were chosen to reject as many background events as
possible without losing too much signal. They were optimized using
real background and simulated signal events by maximizing the so-called 
significance: $S/{\sqrt{N_s+N_b}}$
where $S$ was the yield from a Monte Carlo data set. The numbers of signal 
($N_s$) and background ($N_b$) events inside the
square root were taken from data (all the events within the mass interval 
of 50 MeV/$c^2$ centered at $D_s$ mass value). The cuts are identical for 
the charge conjugate modes.  None of the results presented here is 
sensitive to this optimization
procedure.

 Since the RICH detector does not separate particles with absolute 
certainty, we expect some small amount of misidentification between pion and kaon that 
causes a reflection of $D^{\pm}$ under the $D_s^\pm$ peak (Detailed work 
on the contamination has been reported in Ref~\cite{iori} measurement of $D_s$ lifetime). 
Only resonant ( $\phi\pi$, $K^*K$) channels were considered for this
analysis to reduce the contribution of these reflections significantly.
$D_s^{\pm}$ charm meson decays to $\phi\pi^{\pm}$ were selected by 
starting with candidate $\phi \rightarrow K^-K^+$ decays. The invariant
mass for two well-reconstructed oppositely charged tracks, identified as 
kaons by 
RICH, was calculated. $\phi$ candidates were those pairs whose invariant 
mass
was within $\pm$10 MeV/c$^2$ of the $\phi$ mass (1020 MeV/c$^2$). 
Similarly, those $KK\pi$ combinations that include a $K\pi$ pair
with an invariant mass value within $\pm$35 MeV/c$^2$ of $K^{0*}$ mass 
(892
MeV/c$^2$) were selected as $D_s^{\pm} \rightarrow K^{0*}(892)K^{\pm}$ 
decays. We have obtained 172$\pm$14 $D_s^-$ and 54$\pm$8 $D_s^+$ in 
$\phi\pi$ channel and  174$\pm$14 $D_s^-$ and 71$\pm$12 $D_s^+$ in $K^*K$ 
channel 
for $x_F>$0.15 with the $\Sigma^-$ beam.

\begin{figure}[h]
\centerline{\psfig{figure=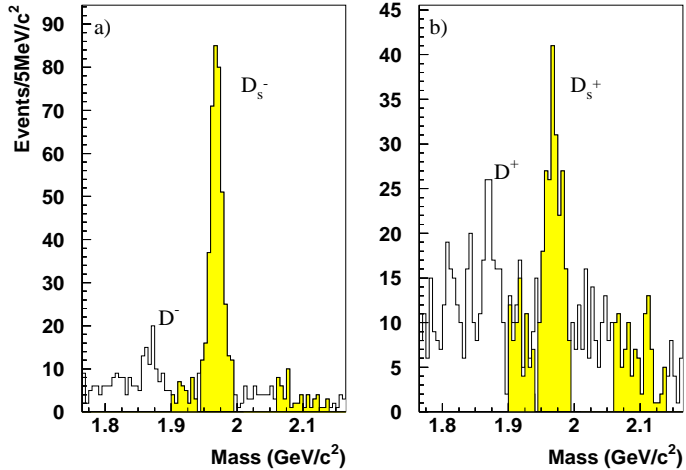,width=90mm}}
\caption{Invariant mass distributions of (a) $KK\pi^-$ and (b) $KK\pi^+$ 
from
the $\Sigma^-$ beam. Mass interval and sidebands used in determining the
yields for asymmetry calculations are shaded in both histograms. The data
shown here include
events having the decays $D_s^{\pm} \rightarrow \phi\pi^{\pm}$ or 
$K^*K^{\pm}$.}
\label{fig:mass}
\end{figure}

With the above cuts the $D_s^\pm$ peaks are clearly evident in the
invariant mass spectra of $KK\pi^-$ and $KK\pi^+$ (Figure~\ref{fig:mass}). 
This figure also shows Cabbibo-suppressed $D^\pm$ peaks. There is a clear
excess of $D_s^-$ over $D_s^+$ as seen in the figure. 

%
%
%
\section{Production Asymmetry and $x_F$ Analysis}

Determining the yields through fitting the mass histograms for a specific
$x_F$ value by a Monte-Carlo generated shape is often inaccurate for
small statistics and fluctuating backgrounds.  In SELEX the yield at a 
specific $x_F$ value was estimated
using a sideband subtraction method assuming a linear background. 
The mass ranges of the sideband background windows were 
[1.900 GeV/$c^2$, 1.940 GeV/$c^2$] and [2.060 GeV/$c^2$, 2.140 
GeV/$c^2$]. We defined asymmetric sidebands to avoid the influence of  
$D^{\pm}\rightarrow K^{+}K^{-}\pi^{\pm}$ and to exclude the D*(2010) mass 
region.

The yields after sideband subtraction  were corrected for the acceptance
(reconstruction efficiency and geometrical acceptance) of the detector.
To estimate the acceptance, $D_s$ events were generated by a 
Monte Carlo program with a flat distribution in
$x_F$ and a Gaussian-distributed transverse momentum with mean $p_t$=0.8
GeV/c. In a given simulation data set the $\rm{D_s}$ decays only into the 
$K^*K$ or $\phi\pi$ mode. Decay tracks were digitized after smearing with 
detector
resolution and multiple Coulomb scattering effects.  The detector hits 
were OR'ed into the hit banks of interaction data. The new hit banks were 
passed through the SELEX off-line software. The acceptance was measured 
using 
the ratio of the reconstructed events to the embedded events. The set of 
cuts that was
used to extract the signal was applied in this case as well. The most
important issue for the asymmetry measurement was the relative efficiency
of $D_s^-$ and $D_s^+$.

\begin{figure}[h]
\centerline{\psfig{figure=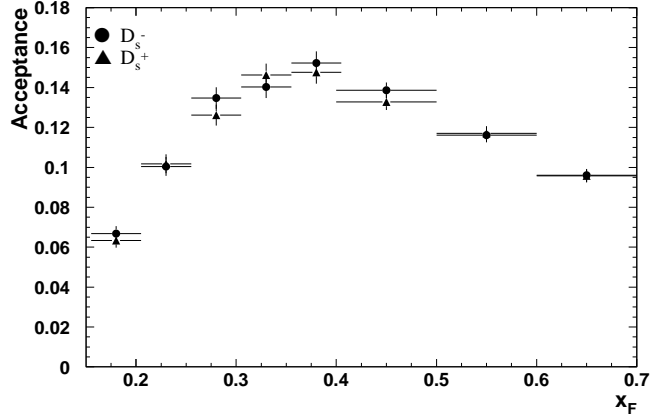,width=90mm}}
\caption{Acceptances for  $D_s^-$ and $D_s^+$ obtained by embedding Monte
Carlo events into data events. The acceptance here is the
combination of geometrical acceptance and reconstruction efficiency of 
the $\phi\pi$ channel.}

\label{fig:eff4paper}
\end{figure}

As indicated in Figure~\ref{fig:eff4paper}, the average difference in
acceptance between $D_s^-$ and $D_s^+$ is very small over all $x_F$ range
compared to the statistical uncertainty. This shows that the spectrometer 
is charge
independent for $D_s$ decay events. In figure only the acceptance for
$\phi\pi$ channel is shown. The acceptance for $K^*K$ channel is slightly 
lower.  The acceptance in $x_F$ is independent of $p_T$.
%
%

\section{Results}

\begin{figure}[h]
\centerline{\psfig{figure=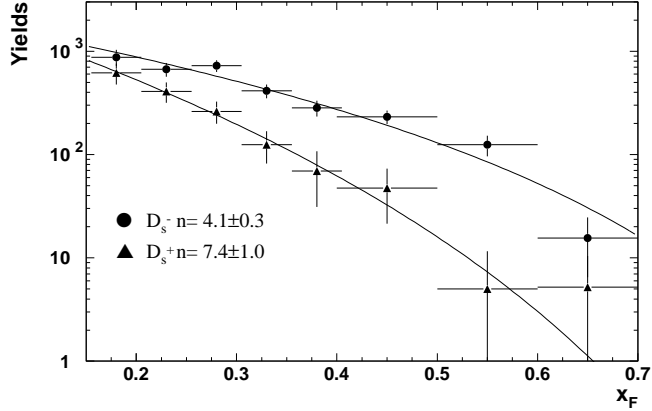,width=90mm}}
\caption{Acceptance-corrected $x_F$ distributions for  $D_s^-$ and 
$D_s^+$ from $\Sigma^-$ beam.
Fits of the yields to $(1-x_F)^n$ for
each charge state are plotted and the $n$-values listed.}
\label{fig:xf}
\end{figure}

After all the cuts and acceptance corrections, the resulting $D_s$ yields
as functions of $x_F$ are shown in Figure~\ref{fig:xf} and 
listed in Table~\ref{xf_asym}. Resulting data points are fit to 
a functional form $(1-x_F)^n$. The values of $n$ obtained from the fits are 
shown on the figure.

Since the beam flux was the same for $D_s^\pm$, the two distributions
compare the relative production differential cross-sections for these 
states. Figure 3 shows that, for the $\Sigma^-$ beam used in this 
measurement, 
$D_s^-$ production is favored over $D_s^+$ at all $x_F$, and the 
difference 
increases at large $x_F$.

\begin{figure}[h]

\centerline{\psfig{figure=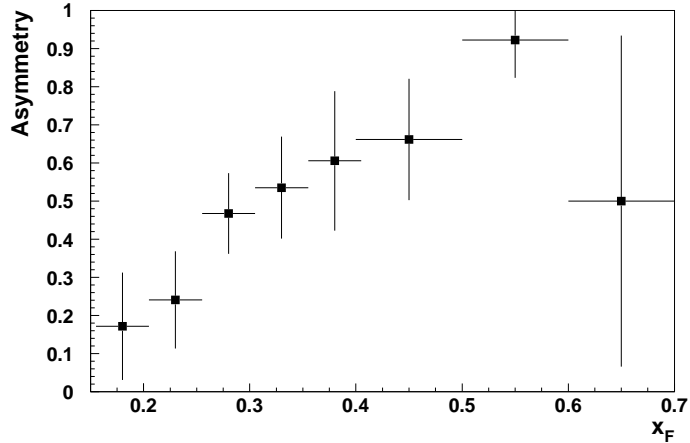,width=90mm}}
\caption{
Production asymmetry for $\Sigma^-$ beam data as a function of $x_F$. 
Yields obtained from resonant ($K^*K$ and $\phi\pi$) events were used to 
calculate the asymmetry.}
 
\label{fig:asym}
\end{figure}

We can discuss this difference in terms of an asymmetry parameter $A$, 
defined as

\begin{eqnarray}
A \equiv {N_{D_s^-} - N_{D_s^+} \over 
  N_{D_s^-} + N_{D_s^+}},
\end{eqnarray}
where $N_{D_s^-}$ and $N_{D_s^+}$ are the corrected yields for $D_s^-$ 
and $D_s^+$, respectively. The asymmetry was calculated for five equally 
divided bins over an $x_F$ range of 0.15 to 0.40 and for three equally 
divided bins over an $x_F$
range of 0.40 to 0.70. Figure~\ref{fig:asym} displays the
acceptance-corrected asymmetry as a function of $x_F$ for the $\Sigma^-$
beam. It shows that there is a significant asymmetry in favor of $D_s^-$
and that the asymmetry increases gradually as $x_F$ increases. The 
asymmetry values are included in Table~\ref{xf_asym}.

\begin{table}

\begin{tabular}{c c c c c}

\\

 $x_F$   & $N_{D_s^-}$       & $N_{D_s^+}$ & Asymmetry\\\hline
0.15-0.20 &875$\pm$156   &619$\pm$141 &0.17$\pm$0.14\\
0.20-0.25 &669$\pm$102    &409$\pm$91  &0.24$\pm$0.13\\
0.25-0.30 &723$\pm$92    &262$\pm$63  &0.47$\pm$0.11\\
0.30-0.35 &413$\pm$62    &125$\pm$43  &0.54$\pm$0.13\\
0.35-0.40 &282$\pm$49    &69$\pm$38   &0.61$\pm$0.18\\
0.40-0.50 &232$\pm$34    &47$\pm$26   &0.66$\pm$0.16\\
0.50-0.60 &124$\pm$28     &5$\pm$7   &0.92$\pm$0.10\\
0.60-0.70 &15$\pm$9      &5$\pm$5     &0.50$\pm$0.43\\

\end{tabular}

\caption{Summary of $D_s^-$ and $D_s^+$ yields and asymmetries from 
$\Sigma^-$ beam. The 
errors are statistical only. Yields are obtained from resonant state 
$K^{*}K$ 
and $\phi \pi$ events.}

\label{xf_asym}
\end{table}

\begin{figure}[h]

\centerline{\psfig{figure=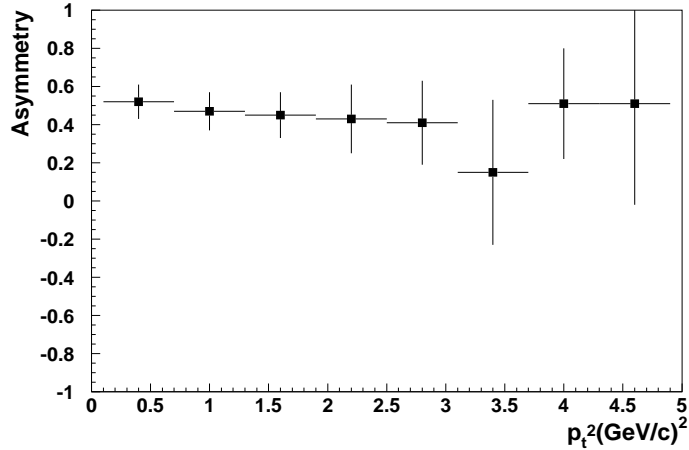,width=90mm}}
\caption{
Production asymmetry for $\Sigma^-$ beam data as a function of $p_t^2$. 
Yields obtained from resonant ($K^*K$ and $\phi\pi$) events were used to 
calculate the asymmetry.}
 
\label{fig:ptasym}
\end{figure}

Figure~\ref{fig:ptasym} displays the asymmetry as a function of $p_t^2$. 
These
asymmetry values are also corrected for acceptance. One can see that the
asymmetry is flat within the observed range (up to $p_t^2 < 5 GeV/c$)

In order to further explore the leading particle effect, the analysis was
repeated with data from the $\pi^-$ beam, obtained under the same
conditions as with the $\Sigma^-$ beam.  
Because pion beam statistics are much lower, we compare only integrated 
asymmetry results for
all $KK\pi$ events with $x_F > 0.15$.  For the $\pi^-$ beam, which has no 
leading
particle, the integrated asymmetry is consistent with zero ($A$ $=$
0.06$\pm$0.11). On the other hand, analysis of the $\Sigma^-$ beam data in 
the same way results in a large asymmetry in favor of $D_s^-$ ($A$ $=$ 
0.53$\pm$0.06).

\section{Systematic Study}

Studies of possible systematic errors due to the side-band subtraction 
method
were done by varying the size and position of the side bands. Effects of 
changing
$x_F$ bin sizes on the results were also investigated. Systematics of the
acceptance calculations have been checked with meson asymmetries and
polarizations, all of which should be zero and are.  The false asymmetry
due to our hardware trigger was also studied. Even when combined together, 
these effects are all considerably smaller than the statistical 
uncertainty and 
are neglected. The contribution of the misidentification of $\pi^-$ beam 
particles as $\Sigma^-$ is estimated to be a few percent.
The resulting dilution in the asymmetry is negligible. 

As mentioned before, background including the $D^{\pm}$ contamination 
under the $D^{\pm}_s$ peak is highly reduced by limiting data to the 
resonant
states. Effects of the remaining background were studied by comparing the 
integrated asymmetries obtained from the two resonant states. 
In the $\phi\pi$ case all backgrounds, including $D^\pm$ contamination, are negligible and cannot affect the asymmetry. The $\phi \pi$ integrated
asymmetry is 0.52$\pm$0.06.  For the $K^*K$ channel it is 0.42$\pm$0.08. 
The effect of the contamination reported in ref~\cite{iori} would reduce
a real $K^*K$ asymmetry of 0.52 to an observed value of 0.48, 
statistically compatible with observation. The overall dilution effect is 
much smaller than the statistical 
uncertainties in individual bins and is not included in the final results 
(Table I and Figure 4).
%
%

\section{Conclusion}

To summarize: the $\Sigma^-(dds)$ beam data show a strong
production asymmetry favoring $D_s^-(\overline c s)$ production.
This is consistent with leading particle effects.  However, the integrated 
asymmetry from $\pi^-$ beam at $x_F>$0.15 for $D_s^\pm$ meson is 
0.06$\pm$0.11, which
is consistent with zero asymmetry as expected since neither $D_s^+$ nor
$D_s^-$ is a leading particle.  Our $\Sigma^-$ results are consistent with 
the previous measurement done by WA89 experiment at CERN with 340 GeV/c 
$\Sigma^-$ beam.

The SELEX pion result of negligible integrated asymmetry agrees with 
the higher-statistics differential distribution for a 500 GeV $\pi^-$ beam 
reported by E791\cite{e791}. Their integrated asymmetry in the $x_F$ range 
0.1 to 0.5 is 0.032$\pm$0.022. Our results also favor the color drag model 
over the intrinsic charm model, since the color drag model predicts a 
large asymmetry at large $x_F$ independent of 
$p_t$\cite{e791,e769,e791_1}.

%
%
%

\section{Acknowledgements}

The authors are indebted to the staff of Fermi National Accelerator 
Laboratory and for invaluable technical support from the staffs of 
collaborating institutions. This project was supported in part by 
Bundesministerium f\"ur Bildung, Wissenschaft, Forschung und Technologie, 
Consejo Nacional de Ciencia y Tecnolog\'{\i}a {\nobreak (CONACyT)}, 
Conselho Nacional de Desenvolvimento Cient\'{\i}fico e Tecnol\'ogico, 
Fondo de Apoyo a la Investigaci\'on (UASLP), Funda\c{c}\~ao de Amparo \`a 
Pesquisa do Estado de S\~ao Paulo (FAPESP), the Israel Science Foundation 
founded by the Israel Academy of Sciences and 
Humanities, Istituto Nazionale di Fisica Nucleare (INFN),
the International Science Foundation (ISF), the National Science 
Foundation (Phy \#9602178), NATO (grant CR6.941058-1360/94), the Russian 
Academy of Science, the Russian Ministry of Science and Technology,
the Turkish Scientific and Technological Research Board (T\"{U}B\.ITAK),
the U.S. Department of Energy (DOE grant DE-FG02-91ER40664 and DOE 
contract
number DE-AC02-76CHO3000), and the U.S.-Israel Binational Science 
Foundation (BSF).

\end{document}